\documentclass[aps,amsmath,showpacs,amsfonts,10pt]{revtex4}
\usepackage{epsfig,graphicx}
\usepackage[english]{babel}
\usepackage{amsfonts}
\usepackage{amsmath}
\usepackage{latexsym}
\usepackage{graphics,bm}
\usepackage{dcolumn}
\usepackage{bm}
\usepackage{rotating}

\begin{document}

\title{Non Degenerate Dual Atomic Parametric Amplifier: Entangled Atomic Fields}

\author{Tarun Kumar$^{1}$ , Aranya B.\ Bhattacherjee$^{2}$ , Priyanka Verma$^{1}$ , Narine Gevorgyan$^{3,4}$ and ManMohan$^{1}$}

\address{$^{1}$Department of Physics and Astrophysics, University of Delhi, Delhi-110007, India} \address{$^{2}$Department of Physics, ARSD College, University of Delhi (South Campus), New Delhi-110021, India}
\address{$^{3}$ Institute for Physical Research, National Academy of Sciences, Ashtarak-2, 0203, Armenia}
\address{$^{4}$ The Abdus Salam International Centre for Theoretical Physics Strada Costiera, 11, I - 34151 Trieste Italy}

\begin{abstract}
In this paper, we investigate the dynamics of two coupled quantum degenerate atomic fields (BEC) interacting with two classical optical fields in the nonlinear atom optics regime. Two photon interaction produces entangled atom-atom pairs which exhibit nonclassical correlations. Since the system involves the creation of two correlated atom pairs, we call it the nondegenerate dual atomic parametric amplifier.
\end{abstract}

\pacs{03.75.-b,03.75.Gg,03.75.Mn}

\maketitle

\section{Introduction}

Ever since Bose Einstein Condensate (BEC) has been achieved in trapped alkali metal atomic vapours \citep{anderson}, it has become a potential and very interesting field because of rich physics involved in it. BEC shows various collective quantum states in Bosonic atoms when trapped in optical lattices. Since atoms in BEC are phase coherent and a large number of bosons occupy the same quantum state, therefor a BEC can be used as an atom Laser \citep{mewes}. Discovery of BEC has also led to a new branch of physics known as non linear quantum optics \citep{lenz} where interaction between light and cold atoms plays a pivotal role.  Foundation of non linear optics has initiated the prediction \citep{goldy}  and observation \citep{stenger} of multiwave mixing in BECs which is basically due to the presence of collisions in them. Interesting quantum phenomena i.e, quantum phase transition: Superfluid to Mott Insulator was also demonstrated using cold atoms \citep{Greiner}.  Light and atoms have different roles to play in optics and atom optics. Various phenomena's which once were supposed to be the properties of light such as, interference, diffraction, beam splitting, reflection etc have also been observed using cold atoms i.e, matter waves. This proves the very importance of study of quantum properties of BEC.
The dynamical interaction between optical and atomic fields gives rise to many interesting optical phenomena. In this paper, we investigate the dynamics of two coupled quantum degenerate atomic field (Bose Einstein Condensate) interacting with two classical optical fields in the nonlinear atom optics regime. Under certain circumstances, one can formally eliminate the dynamics of the atomic field, resulting in effective interactions between light waves. Under a different set of conditions, one can eliminate the electromagnetic field dynamics, resulting in effective atom-atom interactions. These are the regimes of nonlinear optics and nonlinear atom optics respectively. These regimes, therefore, represent limiting cases, where either the atomic or optical field is not dynamically independent, and instead follows the other field in some adiabatic manner which allows for its effective elimination. Two photon interaction produces entangled atom-atom pairs which exhibit nonclassical correlations. Since the system involves the creation of two correlated atom pairs, we call it the nondegenerate dual atomic parameteric amplifier (APA).The propose of this paper is to develop a theory for the dual APA, with an emphasis on the manipulation and control of their quantum statistics and the generation of quantum correlations and entanglement between atomic fields. In this work we will closely follow a previous work done by Moore et al \citep{10}. In particular, we study the Cauchy Schwartz inequality and two mode squeezing. Violation of the Cauchy Schwartz inequality indicates the quantum nature of the correlations. The violation of Cauchy-Schwartz inequality is found to exist for a large number of atoms and is a macroscopic quantum phenomena. The violation of Cauchy-Schwartz inequality has been studied extensively in number of systems like in collective resonance fluorescence \citep{11}, two photon cascade \citep{12} , two photon laser \citep{13} , Parametric amplifier \citep{14}, optical double resonance \citep{15} , in even and odd trio coherent state \citep{16} , four wave mixing \citep{17}. Violation of Cauchy Schwartz inequality has also been observed experimentally, though only in few system \citep{18}.

\section{The Model}
\indent The atoms in the BEC are atomic systems in the ground hyperfine sublevels F=1, M$_F$ = $\pm$1,0,...., labeled respectively as $| g_{\pm},0>$. Two counter propagating $\sigma^+$ and $\sigma^-$ traveling wave light fields are employed to excite the atomic condensate via two photon excitations. Both the light waves are assumed to be strong enough to be treated as classical fields. To initiate our arguments, we begin with the second quantized Hamiltonian

\begin{eqnarray}
H = H_{Atom} + H_{Laser} + H_{Atom-Laser} + H_{Atom-Atom},
\end{eqnarray}

where $H_{Atom}$ gives the free evolution of the atomic fields, $H_{Atom-Laser}$ describes the dipole coupling between the atomic fields and the laser fields. $H_{Atom-Atom}$ contains the two body s-wave scattering collisions between ground-state atoms. With energy of the $|g_->$ state taken to be the reference zero, the total Hamiltonian is given by-

\begin{eqnarray}
H &=& \sum_{i,j=g_{\pm,e}}\int d^3x \psi_{i}^{\dag}(x,t) \Bigg(\frac{-\hbar^2\bigtriangledown^2}{2m}+V_i(x)\Bigg)\psi_i(x,t) \nonumber \\ &+& \int d^3x \psi_{e}^{\dag}(x,t) \hbar \bigtriangleup
\psi_{e}(x,t) \nonumber \\ &-& \frac{\hbar}{2}
\Bigg(\int d^3x \Omega_1\psi_{e}^{\dag}(x,t) e^{ikx}\psi_{g_-}(x,t) +\int d^3x \Omega_2\psi_{e}^{\dag}(x,t) e^{-ikx}\psi_{g_+}(x,t) +H.C.\Bigg) \nonumber \\ &+& \frac{1}{2} \int d^3x k_{ij}\psi_{i}^{\dag}(x,t)\psi_{j}^{\dag}(x,t)\psi_{j}(x,t)\psi_{i}(x,t),
\end{eqnarray}

where $\psi_i$$(x,t)$ and  $\psi_i^{\dag}$$(x,t)$ are the annihilation and creation atomic field operators at position $x$ and in the states $i$ $=$ $g_{\pm}$,$e$ and at time $t$. They obey the usual bosonic equal time commutation relations \Big[$\psi_i$$(x)$,$\psi_j^{\dag}(x^{'})$\Big]=$\delta_{i,j} $ $\delta(x,x^{'})$ and \Big[$\psi_i$$(x)$,$\psi_j(x^{'})$\Big]=\Big[$\psi_i^{\dag}$$(x)$,$\psi_j^{\dag}(x^{'})$\Big]$=$ $0$. $V_i$ is the trap potential, $\bigtriangleup$ is the laser detuning, assumed to be sufficiently large to reduce the incoherent effect of spontaneous emission during the interaction of the laser fields with the condensate. $\Omega_1$and $\Omega_2$ are the Rabi frequencies of the $\sigma^+$ and  $\sigma^-$ laser fields respectively. $k_{ij}$=$\frac {4\pi \hbar a_{ij}}{M}$ , $a_{ij}$ denotes the s-wave scattering lengths for scattering between atoms in the internal states $i$ and $j$ and $M$ is the mass of a single atom. Since the density of the excited state atoms is assumed to be small, we shall neglect two body collisions between excited atoms ($k_{ee}$ $=$ $0$) and that between excited and ground state atoms ($k_{eg_{\pm}}$ $=$ $0$). To simplify the analysis further, we assume $k_{g_{-}g_{-}}$ = $k_{g_{+}g_{+}}$ = $k_{g_{+}g_{-}}$ = $k_{g_{-}g_{+}}$ $=$ $k_{0}$, which is particularly true for rubidium. In terms of the Hamiltonian (2), the atomic field operators satisfy the following Heisenberg equation of motion.

\begin{eqnarray}
i \hbar \frac{d}{dt}\psi_{e(x,t)} = -\Bigg(\hbar \bigtriangleup \psi_{e(x,t)} + \frac{\hbar}{2} \Omega_1\psi{_{g_{-}}} e^{ikx} + \frac{\hbar}{2} \Omega_2\psi{_{g_{+}}} e^{-ikx}\Bigg),
\end{eqnarray}

\begin{eqnarray}
i \hbar \frac{d}{dt}\psi_{g_{{+}}}(x,t) = -\Bigg((\frac{\hbar^{2} \bigtriangledown^{2}}{2m} - V_{g})\psi_{g_{{+}}}(x,t) + \frac{\hbar}{2} \Omega{_{2}}^{*}e^{ikx}\psi_e(x,t)\Bigg) + k_{0}\Bigg(|\psi_{g_{-}}|^2 + |\psi_{g_{+}}|^2\Bigg) \psi_{g_{{+}}}(x,t),
\end{eqnarray}

\begin{eqnarray}
i \hbar \frac{d}{dt}\psi_{g_{{-}}}(x,t) = -\Bigg((\frac{\hbar^{2} \bigtriangledown^{2}}{2m} - V_{g})\psi_{g_{{-}}}(x,t) + \frac{\hbar}{2} \Omega{_{1}}^{*}e^{-ikx}\psi_e(x,t)\Bigg) + k_{0}\Bigg(|\psi_{g_{-}}|^2 + |\psi_{g_{+}}|^2\Bigg) \psi_{g_{{-}}}(x,t),
\end{eqnarray}

where we have dropped the kinetic energy and trap potential in Eqn.(3) under the assumption that the life time of the excited atom, which is of the order of $\frac{1}{\bigtriangleup}$, is so small that the atomic center of mass motion may be safely neglected during this period. For the same reason, we are justified in neglecting collisions between excited atoms. The trap potentials for the ground state $|g_{+}>$ and $|g_{-}>$ are assumed to be similar. We allow the excited atoms to adiabatically follow the motion of the ground state atoms. In this case Eqn.(3) has the adiabatic solution

\begin{eqnarray}
\psi_e(x,t) = -\frac{1}{2 \bigtriangleup}\Bigg(\Omega_{1}e^{ikx}\psi_{g_{-}}(x,t) +\Omega_{2}e^{-ikx}\psi_{g_{+}}(x,t)\Bigg).
\end{eqnarray}

Substituting Eqn.6 into the equation of motion for $\psi_{g_{\pm}}(x,t)$, we arrive at the effective Heisenberg equation of motion for the ground state field operators.

\begin{eqnarray}
i\frac{d}{dt}\psi_{g_{+}}(x,t) = \Bigg(\zeta + \frac{|\Omega_2|^{2}}{4\bigtriangleup} + k_{0}^{'}N\Bigg) \psi_{g_{+}}(x,t) + \frac{\Omega_{1}\Omega_{2}^*}{4\bigtriangleup}e^{i2kx}\psi_{g_{-}}(x,t)
\end{eqnarray}

\begin{eqnarray}
i\frac{d}{dt}\psi_{g_{-}}(x,t) = \Bigg(\zeta + \frac{|\Omega_1|^{2}}{4\bigtriangleup} + k_{0}^{'}N\Bigg)\psi_{g_{-}}(x,t) + \frac{\Omega_{1}^{*}\Omega_{2}}{4\bigtriangleup}e^{-i2kx}\psi_{g_{+}}(x,t),
\end{eqnarray}

where $k_{0}^{'}$ = $\frac{k_{0}}{2 \hbar}$, $\zeta$ = $\frac{-\hbar \bigtriangledown^{2}}{2m}$ + $\frac{V_{g}}{\hbar}$. The atom number conservation holds if the number of ultra cold atoms are large enough so that the small fluctuations in the total population are neglected. Consequently $|\psi_{g_{+}}|^{2} +|\psi_{g_{-}}|^{2}=N$. Eqn.(7) and Eqn.(8) show that the two photon process transfers the atoms from the condensate ground states $|g_{+}>$ and $|g_{-}>$to new states that are shifted in momentum space by the two-photon recoil. These new states constitutes secondary condensate components, which can be considered as momentum side modes to the original condensates. As these sides modes are populated, they begin to interfere with the original condensates, resulting in fringe patterns.

We now introduce the atomic field operators $C_{g_{\pm}}$ which annihilates an atom in the condensate ground state $|\psi_{g_{+}}>$,

\begin{eqnarray}
C_{g_{\pm}} = \int d^{3}x \psi_{0}^{*}\psi_{g_{\pm}},
\end{eqnarray}

where $\psi_0$ satisfies the time-independent Gross-Pitaevskii equation

\begin{eqnarray}
 (\zeta + 2k_{0}^{'}N|\psi_0|^{2} - \frac{\mu}{\hbar})\psi_{0}(x)=0.
\end{eqnarray}

Here $\mu$ being the chemical potential. Differentiating Eqn.(9) with respect to time and inserting Eqns. (7), (8) and (10), we find the equation of motion for $C_{g_{\pm}}$.

\begin{eqnarray}
 \frac{d}{dt}C_{g_{+}} = -i\int d^3x \Bigg(\Big[\frac{\mu}{\hbar} + k_{0}^{'}N - 2k_{0}^{'}N|\psi_{0}|^{2} + \frac{|\Omega_{2}|^{2}}{4\bigtriangleup}\Big]\psi_{0}^{*} \psi_{g_{+}} + \frac{\Omega_{1}\Omega_{2}^{*}}{4\bigtriangleup}e^{2ikx}\psi_{g_{-}}\psi_{0}^{*}\Bigg),
\end{eqnarray}

\begin{eqnarray}
 \frac{d}{dt}C_{g_{-}} = -i\int d^3x \Bigg(\Big[\frac{\mu}{\hbar} + k_{0}^{'}N - 2k_{0}^{'}N|\psi_{0}|^2 + \frac{|\Omega_{1}|^{2}}{4\bigtriangleup}\Big]\psi_{0}^{*} \psi_{g_{-}} + \frac{\Omega_{1}^{*}\Omega_{2}}{4\bigtriangleup}e^{-2ikx}\psi_{g_{+}}\psi_0^{*}\Bigg).
\end{eqnarray}

From Eqn.(11) and Eqn.(12), we see that the effect of the optical fields is to couple the original condensate modes to two side modes (whose momentum differ by 2$\hbar$k from the original modes). The operators for the side modes are defined as-

\begin{eqnarray}
 C_{g_{--}} = \int d^{3}x \psi_{0}^{*}e^{2ikx}\psi_{g_{-}},
\end{eqnarray}

\begin{eqnarray}
 C_{g_{++}} = \int d^{3}x \psi_{0}^{*}e^{-2ikx}\psi_{g_{+}}.
\end{eqnarray}

Eqn.(11) and Eqn.(12) are rewritten in terms of Eqn.(13) and Eqn.(14) as

\begin{eqnarray}
 \frac{d}{dt}C_{g_{+}} = -i\alpha C_{g_{+}} - i\beta C_{g_{--}},
\end{eqnarray}

\begin{eqnarray}
 \frac{d}{dt}C_{g_{-}} = -i\gamma C_{g_{-}} - i\delta C_{g_{++}},
\end{eqnarray}

where,

\begin{eqnarray}
 \alpha &=& {\frac{\mu}{\hbar} + k_{0}^{'}N - 2k_{0}^{'}N|\psi_{0}|^2 + \frac{|\Omega_{2}|^{2}}{4\bigtriangleup}}, \\ \gamma &=& {\frac{\mu}{\hbar} + k_{0}^{'}N - 2k_{0}^{'}N|\psi_{0}|^{2} + \frac{|\Omega_{1}|^{2}}{4\bigtriangleup}}, \\ \beta &=& \frac{\Omega_1\Omega_{2}^{*}}{4\bigtriangleup},\\
 \delta &=& \frac{\Omega_{1}^{*}\Omega_{2}}{4\bigtriangleup}.
\end{eqnarray}

Eqns.(15) and (16) indicate the coupling between the modes $C_{g_{+}}$,($C_{g_{-}}$) and $C_{g_{--}}$ ($C_{g_{++}}$). We now derive the Heisenberg equations for the momentum side mode field operators $C_{g_{--}}$,$C_{g_{++}}$. Differentiating Eqns. (13) and (14) with respect to time and again inserting Eqns.(7) and (8), we get

\begin{eqnarray}
 \frac{d}{dt}C_{g_{++}} = -i\alpha C_{g_{++}} - i\beta C_{g_{-}},
\end{eqnarray}

\begin{eqnarray}
 \frac{d}{dt}C_{g_{--}} = -i\gamma C_{g_{--}} - i\delta C_{g_{+}}.
\end{eqnarray}

The operators $C_{i}$ ($i=g_{+},g_{-},g_{++},g_{--}$) obey the commutation relations \Big[$C_{i}$,$C_{j}^{\dag}$\Big] = $\delta_{ij}$, all other commutators are zero. The set of Eqns. (15),(16),(21) and (22) represent coupled equations. For real $\Omega_{1}$ and $\Omega_{2}$, the solutions are obtained as.

\begin{eqnarray}
 C_{g_{+}}(t) = [C_{g_{+}}(0)\delta_{1} - C_{g_{--}}(0)\delta_{2}]e^{-iR^{'}t},
\end{eqnarray}

\begin{eqnarray}
 C_{g_{-}}(t) = [C_{g_{-}}(0)\delta_{1} - C_{g_{++}}(0)\delta_{2}]e^{-iR^{'}t},
\end{eqnarray}

\begin{eqnarray}
 C_{g_{--}}(t) = [C_{g_{--}}(0)\delta_{1} - C_{g_{+}}(0)\delta_{2}]e^{-iR^{'}t},
\end{eqnarray}

\begin{eqnarray}
 C_{g_{++}}(t) = [C_{g_{++}}(0)\delta_{1} - C_{g_{-}}(0)\delta_{2}]e^{-iR^{'}t},
\end{eqnarray}

where

\begin{eqnarray}
\delta_{1} &=& \frac{\Omega_{1}^{2}e^{-iRt} + \Omega_{2}^{2}}{\Omega_{1}^{2} + \Omega_{2}^{2}}, \\
\delta_{2} &=& \frac{\Omega_{1}\Omega_{2}(e^{-iRt}-1)}{\Omega_{1}^{2} + \Omega_{2}^{2}},\\
R^{'} &=&\frac{\mu}{\hbar} + k^{'}_{0}N - 2k^{'}_{0}N|\psi_{0}|^2 ,\\
R &=& \frac{\Omega_{1}^{2} + \Omega_{1}^{2}}{4\bigtriangleup}.
\end{eqnarray}

\begin{figure}[h]
\hspace{-0.0cm}
\includegraphics [scale=0.8]{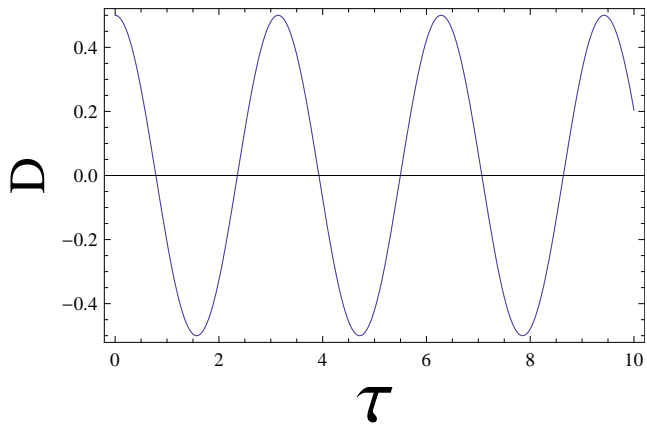}
\caption{Plot of the population difference $D=\frac{|C_{g+}|^{2}-|C_{g++}|^{2}}{N}$ as a function of dimensionless time $\tau=\Omega t$ for $\frac{\Omega}{4 \Delta}=0.1$.}
\label{1}
\end{figure}

In order to understand the physics implied by Eqns. (23) - (26), we take for simplicity $\Omega_{1}=\Omega_{2} = \Omega$ and initially the original modes are equally populated i.e, $C_{g_{-}}(0) = C_{g_{+}}(0) = \sqrt{\frac{N}{2}}$ and all other higher modes are empty $C_{g_{++}}(0)$ $=$ $C_{g_{--}}(0)$ $=$ $0$.
Consequently, the Eqns 23 - 26 reduces to

\begin{eqnarray}
 C_{g_{+}}(t) = C_{g_{-}}(t) = \sqrt{\frac{N}{2}} \cos(\frac{\Omega^{2}t}{4\bigtriangleup})e^{-iR^{'}t},
\end{eqnarray}

\begin{eqnarray}
 C_{g_{++}}(t) = C_{g_{--}}(t) = -i\sqrt{\frac{N}{2}} \sin(\frac{\Omega^{2}t}{4\bigtriangleup})e^{-iR^{'}t}.
\end{eqnarray}

The total atomic number density is always conserved,

\begin{eqnarray}
 |C_{g_{+}}(t)|^2 + |C_{g_{-}}(t)|^2 + |C_{g_{++}}(t)|^2 + |C_{g_{--}}(t)|^2 = N
\end{eqnarray}

The atomic field operators $C_{g_{++}}^{\dagger}C_{g_{-}}$ and  $C_{g_{--}}^{\dagger}C_{g_{+}}$ correspond physically to interference fringes that appear because of periodic modulation of the atomic density. In figure 1, we plot the population difference $D=\frac{|C_{g+}|^{2}-|C_{g++}|^{2}}{N}$ as a function of dimensionless time $\tau=\Omega t$. The observed population oscillations are in fact Rabi cycling of the atoms between the original condensates and the side modes. Eqn.(31) and (32) indicate that the population transfer between the modes can be conveniently manipulated by the laser parameters.

\section{Quantum Statistics}

In this section, we use solutions 25 and 26 to calculate some of the quantum statistical properties of the  system. One of the most important application of the present dual APA system is the generation of entangled atomic states. We first examine the second order two mode intensity correlation functions which give measure of entanglement and can be used to determine whether or not non classical correlation exist between the various modes. We then discuss two mode squeezing in the present system

\begin{figure}[t]

\begin{tabular}{cc}
\includegraphics [scale=0.5] {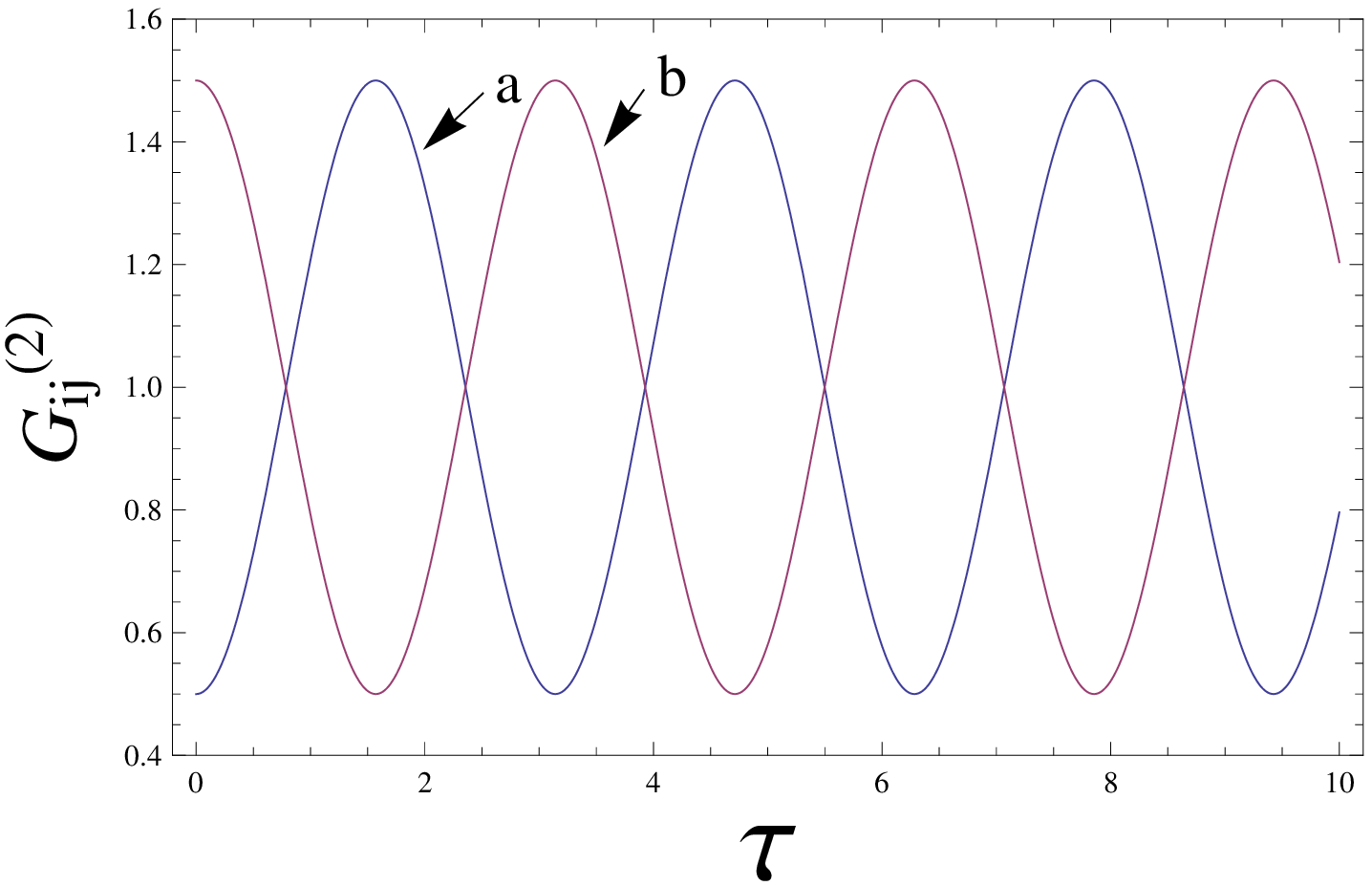}&\includegraphics [scale=0.5] {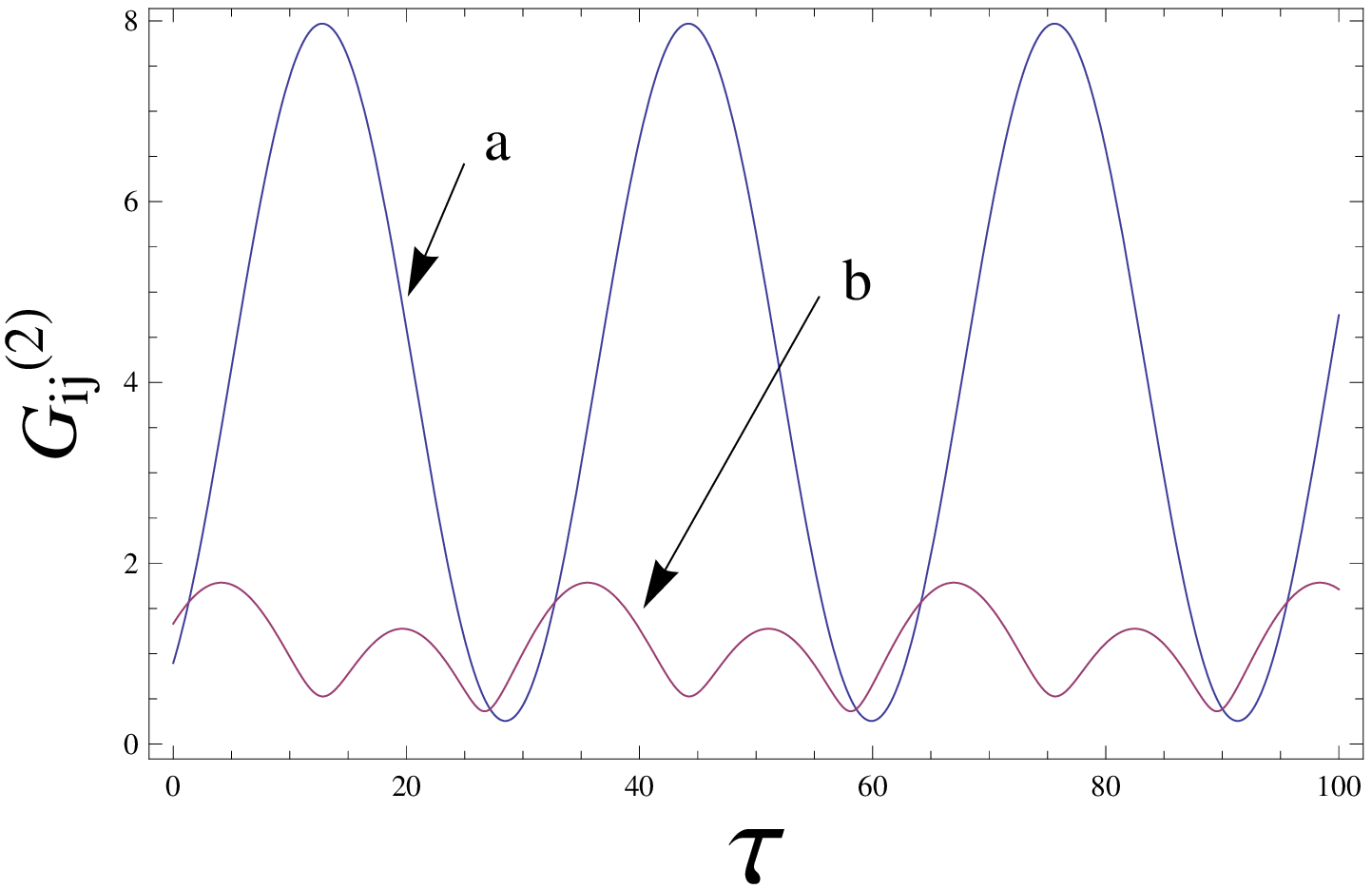}\\
\end{tabular}

\caption{Figure illustrating the violation of Cauchy Schwartz inequality. Plot (a) is $G^{(2)}_{g_{++}g_{-}}(\tau)$ and (b) is $[G_{g_{++}g_{++}}]^{1/2}[G_{g_{-}g_{-}}]^{1/2}$. The left plot is for $\frac{\Omega}{4 \Delta}=1.0$ and the right plot is for $\frac{\Omega}{4 \Delta}=0.1$. }

\label{figure2}
\end{figure}

\subsection{Atom-Atom entanglement}

The second order correlation function of the various condensate modes have the following time average form

\begin{eqnarray}
G^{(2)}_{ij}(\tau) = \frac{<N_{i}N_{j}> - \delta_{ij}<N_{j}>}{<N_{i}><N_{j}>}
\end{eqnarray}

where, $N_i$ $=$ $C^{\dag}_{i}(t)C_{i}(t+\tau)$

The two mode correlation function ($i\neq j$) arise, e.g., if we consider a measurement of the intensity difference between two modes, described by the operators $N_{ij}$ = $N_{i}$-$N_{j}$. For uncorrelated fields, $G_{ij}^{(2)} = 1$. If, however, there are correlations between fluctuations in the intensities of the two modes, then we have $G_{ij}^{(2)} > 1$. For classical fields, the two modes ($i\neq j$ ) correlations are constrained by the Cauchy Schwartz inequality

\begin{eqnarray}
G_{ij}^{(2)} \leq [G_{ii}]^{1/2}[G_{jj}]^{1/2}
\end{eqnarray}

This inequality indicates that for a classical system the cross correlation between the two atomic fields $G_{ij}$ cannot be larger than the geometric mean of the zero time autocorrelations $G_{ii}$ and $G_{jj}$. Quantum mechanical fields, however, can violate this inequality. Fig. 2 illustrate the time dependence of the correlation functions  $G^{(2)}_{g_{++}g_{-}}(\tau)$ and $[G_{g_{++}g_{++}}]^{1/2}[G_{g_{-}g_{-}}]^{1/2}$ for $\frac{\Omega}{4\bigtriangleup} = 1.0$ (left plot) and $0.1$ (right plot) respectively. Results indicate the sensitivity of the statistical properties of the BEC to the parameters  $\frac{\Omega}{4\bigtriangleup}$. When $\frac{\Omega}{4\bigtriangleup} = 1.0$,  Cauchy Schwartz inequality is violated for certain times periodically. Violation of the Cauchy Schwartz inequality for almost all times is seen for $\frac{\Omega}{4\bigtriangleup} = 0.1$.  The Cauchy-Schwartz inequality is violated since the correlation between the atomic fields of different modes is larger than the correlation of atomic fields of the same mode. The violation of the Cauchy Schwartz inequality represents a non-classical effect. Strong correlations between two different modes make them indistinguishable, a key requirement for quantum information processing protocols and form the basis of proposed quantum memories and logic.   Violation of this inequality is also an indication of correlation between the fluctuation in the intensities of the two modes $C_{g_{++}}$ and $C_{g_{-}}$. For a given value of $\frac{\Omega}{4\bigtriangleup}$, this violation is restricted to only a certain range of the time quadrature. This range can be easily adjusted by the parameter $\frac{\Omega}{4\bigtriangleup}$.

\begin{figure}[t]

\begin{tabular}{c}
\includegraphics [scale=0.5] {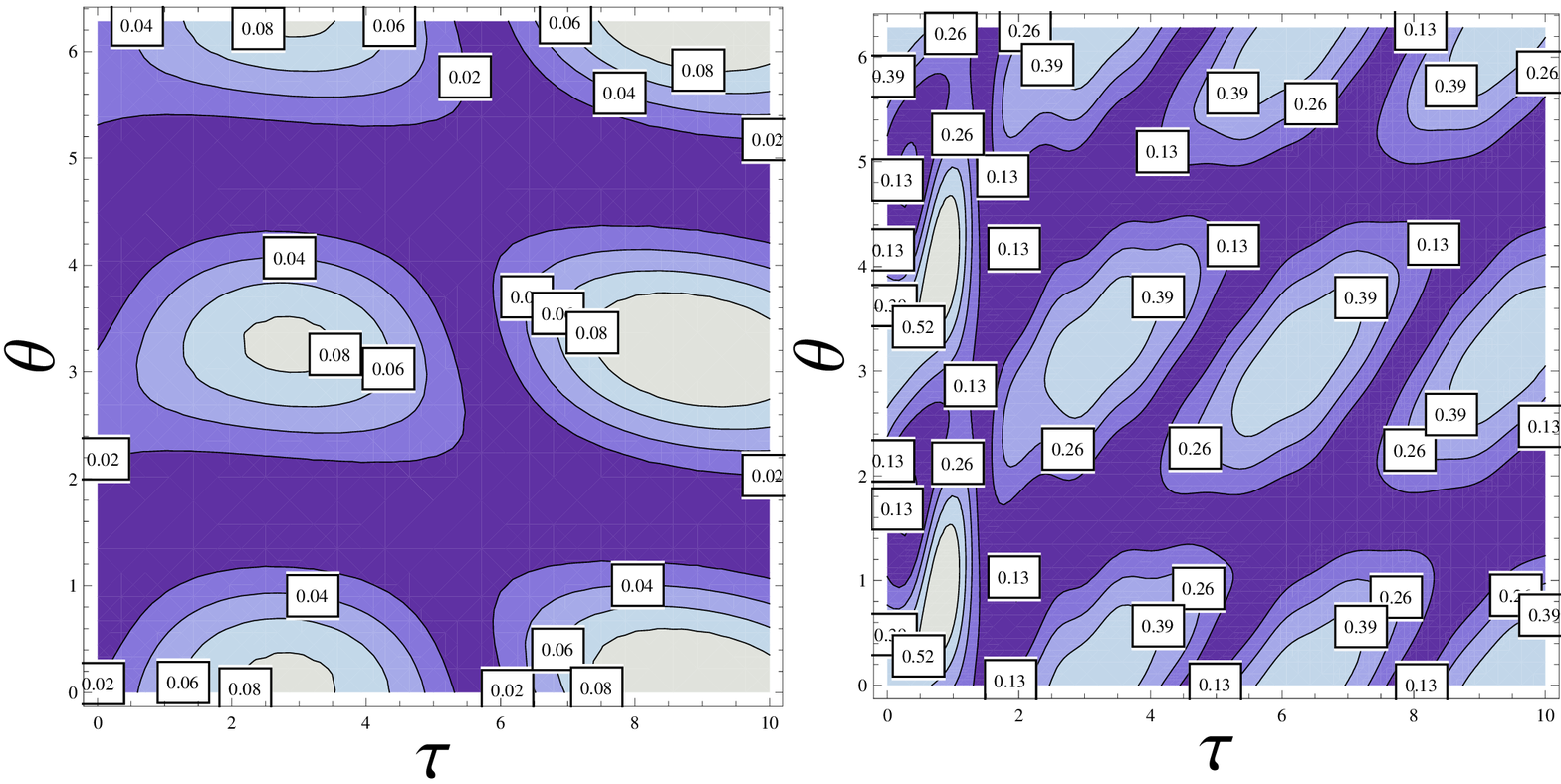}\\ \includegraphics [scale=0.5] {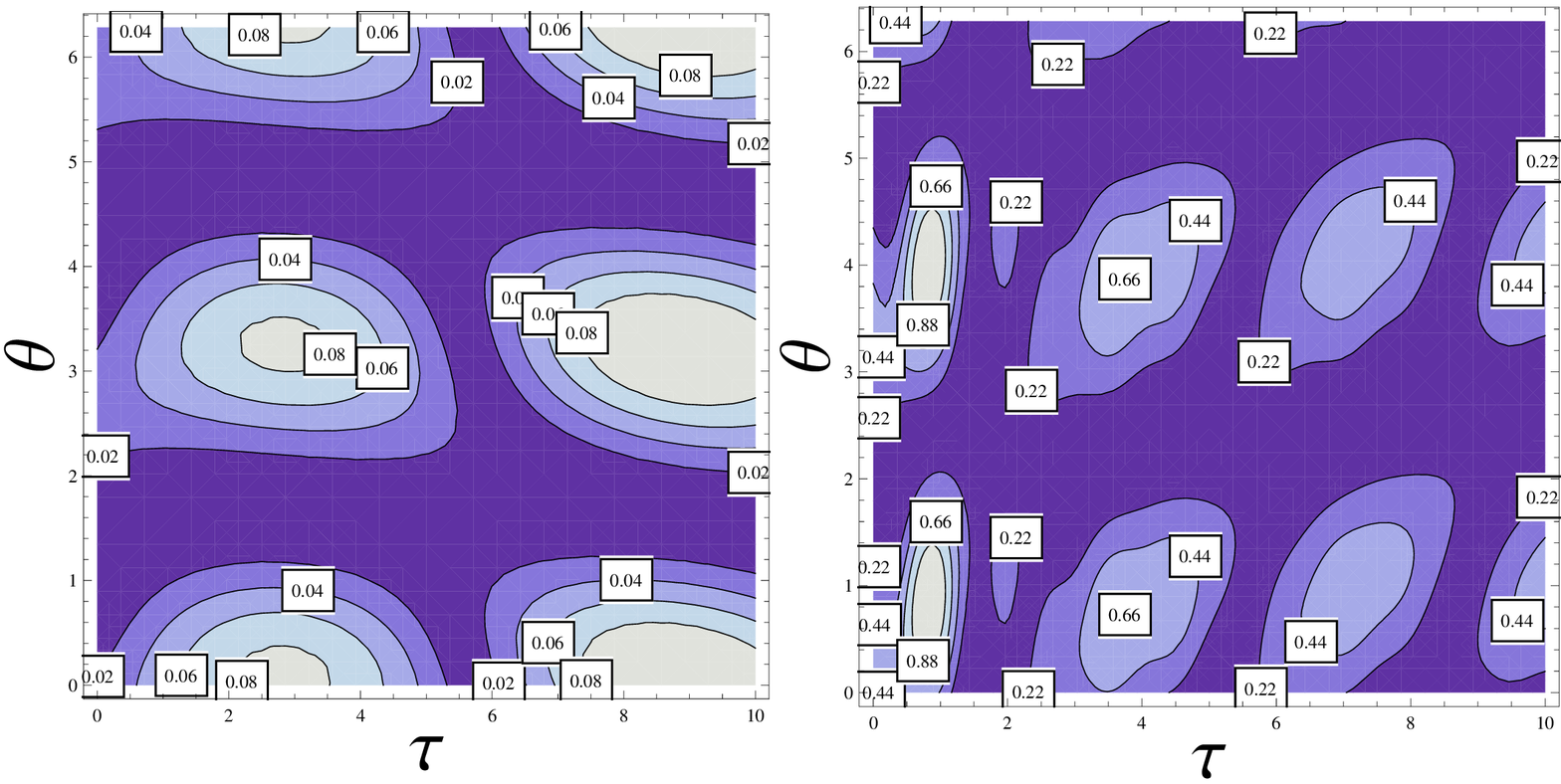}
 \end{tabular}

\caption{Contour plot of $V(X_{\theta})$ as a function of time $\tau$ and $\theta$. The parameters are $\Omega / 4\Delta = 0.1$ and $R' = 0$ (top left plot), $\Omega / 4\Delta = 1.0$ and $R' = 0$ (top right plot), $\Omega / 4\Delta = 1.0$ and $R' = 0.1$ (bottom left plot), $\Omega / 4\Delta = 0.1$ and $R' = 0.1$ (bottom right plot).
 }

\label{figure3}
\end{figure}

\subsection{Two mode squeezing}
\indent In this subsection, we discuss the phase sensitive two mode correlation in the dual APA, which we expect to show squeezing behaviour. The quantum correlations discussed previously, may give rise to two mode squeezing. The APA exhibits quantum mechanical correlations which violate the Cauchy Schwartz inequality. These quantum correlations may be further exploited to give squeezing. In particular, we study the squeezing properties of the two modes C$_{g{-}}$ and  C$_{g{++}}$ as in the previous subsection. We introduce the quadrature components of the superposed modes  C$_{g{-}}$ and  C$_{g{++}}$
\begin{eqnarray}
X_{\theta} = \sqrt{\frac{N}{2}}\Bigg( C_{g{-}}e^{i\theta} +  C_{g{++}}e^{-i\theta} + H.C.\Bigg)
\end{eqnarray}

The variance of X$_{\theta}$ is given by-

\begin{eqnarray}
V(X_{\theta}) = <X_{\theta}^{2}> - <X_{\theta}>^{2},
\end{eqnarray}

where $\theta$ is the squeezing parameter and can be varied to minimize the quadrature variance (V$_{min}$). The Heisenberg uncertainity principal gives V($X_{\theta}$)V($X_{\theta+ \frac{\pi}{2}}$) $\geq$ 1; hence a quadrature component is squeezed, provided  V($X_{\theta}$) $\leq$1. Fig. 3 shows the contour plot of $V(X_{\theta}$)as a function of time and $\theta $. Changing the squeezing parameter $\theta$, enables one to move from enhanced to diminished, i.e. squeezed, quadrature phase fluctuations. The system parameters $\Omega/4\Delta$ and $R^{'}$ coherently controls the squeezing. Note that for a fixed chemical potential and fixed particle number, $R^{'}$ can be tuned by the two body interaction. Appropriate choice of the system parameters leads to persistent squeezing (squeezing to exist for all times). This is similar to the two-mode optical parametric amplifier, where the quadrature component remains squeezed for all times. We note that the squeezing in the APA is due to the quantum correlations which build up in the BEC original modes and the side modes. The individual modes are not squeezed. However it has been shown that the presence of squeezing does not necessarily guarantee violation of the Cauchy-Schwartz inequality \citep{18}.

\section{Conclusions}

In this work, we have investigated the quantum dynamics of two coupled quantum degenerate atomic fields (BEC) interacting with two classical optical fields in the nonlinear atom optics regime. Two photon interaction produces entangled atom-atom pairs which exhibit nonclassical correlations. We have shown that by tuning the various system parameters, we can coherently control the two-mode squeezing and the Cauchy-Schwartz inequality. We can go into the strong quantum regime by regulating the system parameters. The violation of the Cauchy-Schwartz inequality and the two-mode squeezing is a direct indication of the system entering the quantum regime. Indistinguishability of the coupled atomic modes in the quantum regime is the key ingredient for possible applications. The current system in the quantum regime is potentially useful platform for quantum information processing protocols and for possible quantum memories. Since the system involves the creation of two correlated atom pairs, we call it the nondegenerate dual atomic parametric amplifier.

\section{Acknowledgements}

The authors Tarun Kumar and Priyanka Verma thanks the University Grants Commission, New Delhi for the  Research Fellowship. Narine Gevorgyan thanks the Abdus Salam ICTP, Trieste, Italy for the facilities under the Junior Associateship Programme. Part of this work was done at the Laboratory of Information Technologies (LIT) of Joint Institute for Nuclear Research, Dubna, Russia.  Narine Gevorgyan would like to thank Director of LIT Prof. Victor Ivanov for the hospitality and financial support.

\end{document}